\documentclass[baaa]{baaa}

\usepackage[pdftex]{hyperref}
\usepackage{subfigure}
\usepackage{natbib}
\usepackage{helvet,soul}
\usepackage[font=small]{caption}

\begin{document}


\journalvol{58}
\journalyear{2015}
\journaleditors{P. Benaglia, D.D. Carpintero, R. Gamen \& M. Lares}


\contriblanguage{1}


\contribtype{1}

\thematicarea{6}

\title{Studying the molecular gas towards the R Coronae Australis dark cloud}
\subtitle{}


\titlerunning{R Coronae Australis dark cloud}


\author{S. Paron\inst{1,2}, M. Celis Peña\inst{1},  M. E. Ortega\inst{1}, M. Cunningham\inst{3},
P. A. Jones\inst{3} \& M. Rubio\inst{4}}
\authorrunning{Paron et al.}


\contact{sparon@iafe.uba.ar}

\institute{Instituto de Astronomía y Física del Espacio (CONICET-UBA) \and
  FADU y CBC, Universidad de Buenos Aires \and
  School of Physics, University of New South Wales, Sydney, NSW 2052, Australia \and
  Departamento de Astronomía, Universidad de Chile, Casilla 36-D, Santiago, Chile 
}


\resumen{
La nube oscura R Coronae Australis es una de las regiones de formación estelar más cercanas al Sol. Ciertamente se sabe
que dicha nube es muy activa en formación estelar conteniendo una gran cantidad de objetos
Herbig-Haro (HHs) y objetos de emisión de línea del hidrógeno molecular (MHOs). 
En este trabajo presentamos los resultados del análisis de observaciones moleculares 
(un mapa de 5.5$^{'}\times5.5^{'}$ de $^{12}$CO J=3--2 y HCO$^{+}$ J=4--3, y un espectro único de N$_{2}$H$^{+}$ J=4--3) 
obtenidas con el telescopio Atacama Submillimeter Telescope Experiment (ASTE) hacia la nube oscura R CrA con una resolución angular y espectral de 
22$^{''}$ y 0.11 km s$^{-1}$, respectivamente. Analizando los espectros de la emisión del  
$^{12}$CO J=3--2 encontramos evidencia cinemática que sugiere la presencia de chorros moleculares
hacia una región poblada por varios objetos HHs y MHOs. La mayoría de dichos objetos se encuentran localizados sobre
el máximo del HCO$^{+}$, sugiriendo que su emisión proviene del incremento de su abundacia debido a la química generada 
por los chorros moleculares.
De manera adicional, presentamos un espectro de N$_{2}$H$^{+}$, siendo la primer detección reportada de esta molécula en la línea
J=4--3 hacia la nube oscura R CrA.
}

\abstract{
The R Coronae Australis dark cloud is one of the closest star-forming
regions to the Sun. The cloud
is known to be very active in star formation, harboring many Herbig-Haro objects (HHs) and
Molecular Hydrogen emission-line Objects (MHOs).
In this work we present results from molecular observations
(a 5.5$^{'}\times5.5^{'}$ map of $^{12}$CO J=3--2 and HCO$^{+}$ J=4--3, and a single spectrum of N$_{2}$H$^{+}$ J=4--3) obtained with the
Atacama Submillimeter Telescope Experiment (ASTE) towards the R CrA dark cloud with an angular and spectral resolution of
22$^{''}$ and 0.11 km s$^{-1}$, respectively. From the $^{12}$CO J=3--2 line we found kinematical spectral
features strongly suggesting the presence of outflows towards a region populated by several HHs and
MHOs. Moreover, most of these objects
lie within an HCO$^{+}$ maximum, suggesting that its emission arises from an increasement of its abundance due to the chemistry triggered by
the outflow activity. Additionally, we are presenting the first reported detection of N$_{2}$H$^{+}$ in the J=4--3 line towards
the R CrA dark cloud.
}



\maketitle

\section{Introduction}
\label{S_intro}

\noindent 
The R Coronae Australis dark cloud (R CrA dark cloud), located at about 130 pc \citep{neuha08}, 
is one of the 
closest star-forming regions to the Sun. The cloud is centered on the 
Herbig Ae/Be stars R CrA and T CrA, which are the exciting sources of the reflection nebula NGC 6729. 
It is known that this dark cloud is very active in star formation (e.g. \citealt{linde12}), 
harboring many Herbig-Haro (HHs) objects and
Molecular Hydrogen emission-line Objects (MHOs). Due to its proximity, this region is
a very interesting target to study the molecular gas related to the star-forming
processes.

\section{Observations}
\label{S_obs}

A region of 5.5$^{\prime}\times5.5^{\prime}$~in size centered at 19:01:51, $-36$:58:20 (J2000) in the 
R CrA dark cloud was mapped in the $^{12}$CO J=3--2 and HCO$^{+}$ J=4--3 lines using 
the Atacama Submillimeter Telescope Experiment (ASTE). The observations were carried out
in August 2010 and were performed in on-the-fly mapping mode achieving an angular sampling of 
6$^{\prime\prime}$. In addition, the N$_{2}$H$^{+}$ J=4--3 line was observed as a single point
at 19:01:52.7, $-36$:57:49 (J2000).
The angular and spectral resolution are 22$^{\prime\prime}$~and 0.11 km~s$^{-1}$, respectively. 
The data were reduced with NEWSTAR\footnote{Reduction software based on AIPS developed at NRAO,
extended to treat single dish data with a graphical user interface (GUI).} and the spectra processed using the XSpec 
software package\footnote{XSpec is a spectral line reduction package for astronomy which has been developed by Per Bergman 
at Onsala Space Observatory.}. Polynomials between first and third order were used for baseline fitting.

\section{Results}
\label{S_res}

\begin{figure}[!ht]
  \centering
  \includegraphics[width=0.5\textwidth]{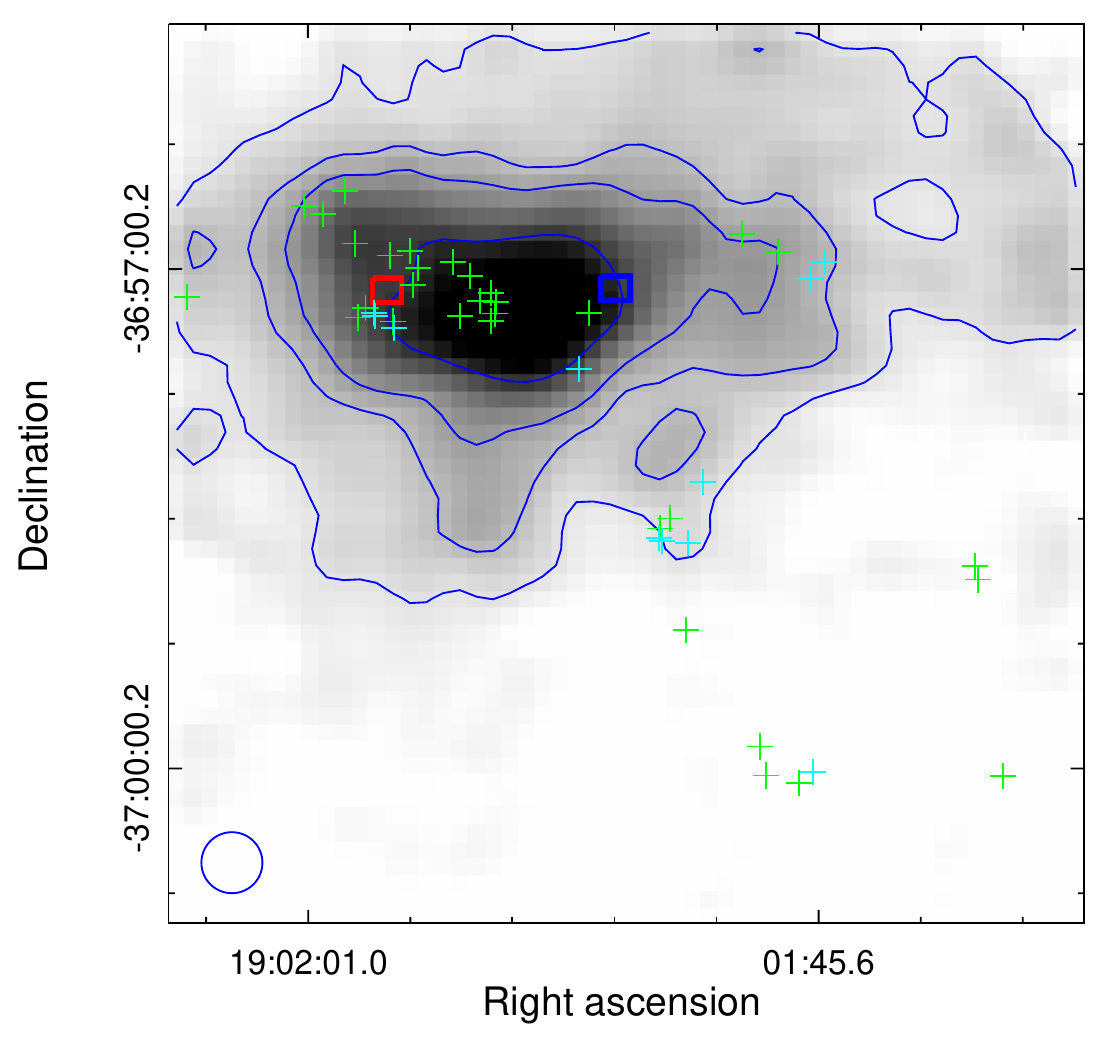}
  \caption{$^{12}$CO J=3--2 emission integrated between $-15$ and $+25$ km~s$^{-1}$. The contour
levels are 30, 50, 70, and 130 K km~s$^{-1}$. The green and cyan crosses represent the positions
of MHO and HH objects, respectively. The red and blue squares are
the positions from which we extracted the spectra presented in Fig.\,\ref{spectra}. The beam is
presented in the bottom-left corner.}
  \label{figinteg}
\end{figure}

Figure\,\ref{figinteg} shows the $^{12}$CO J=3--2 emission integrated between $-15$ and
$+25$ km~s$^{-1}$ towards the R CrA dark cloud.
In addition to the molecular study, we search for HHs objects and MHOs lying in the region from the catalogues
of \citet{reip00} and \citet{davis10}. This kind of objects are strong evidences of active star formation.
The crosses in Fig.\,\ref{figinteg} show the position of HHs and MHOs, which lie
mainly within the molecular gas emission, strongly suggesting that they are embedded in the
molecular cloud.

\begin{figure}[!ht]
  \centering
  \includegraphics[width=0.5\textwidth]{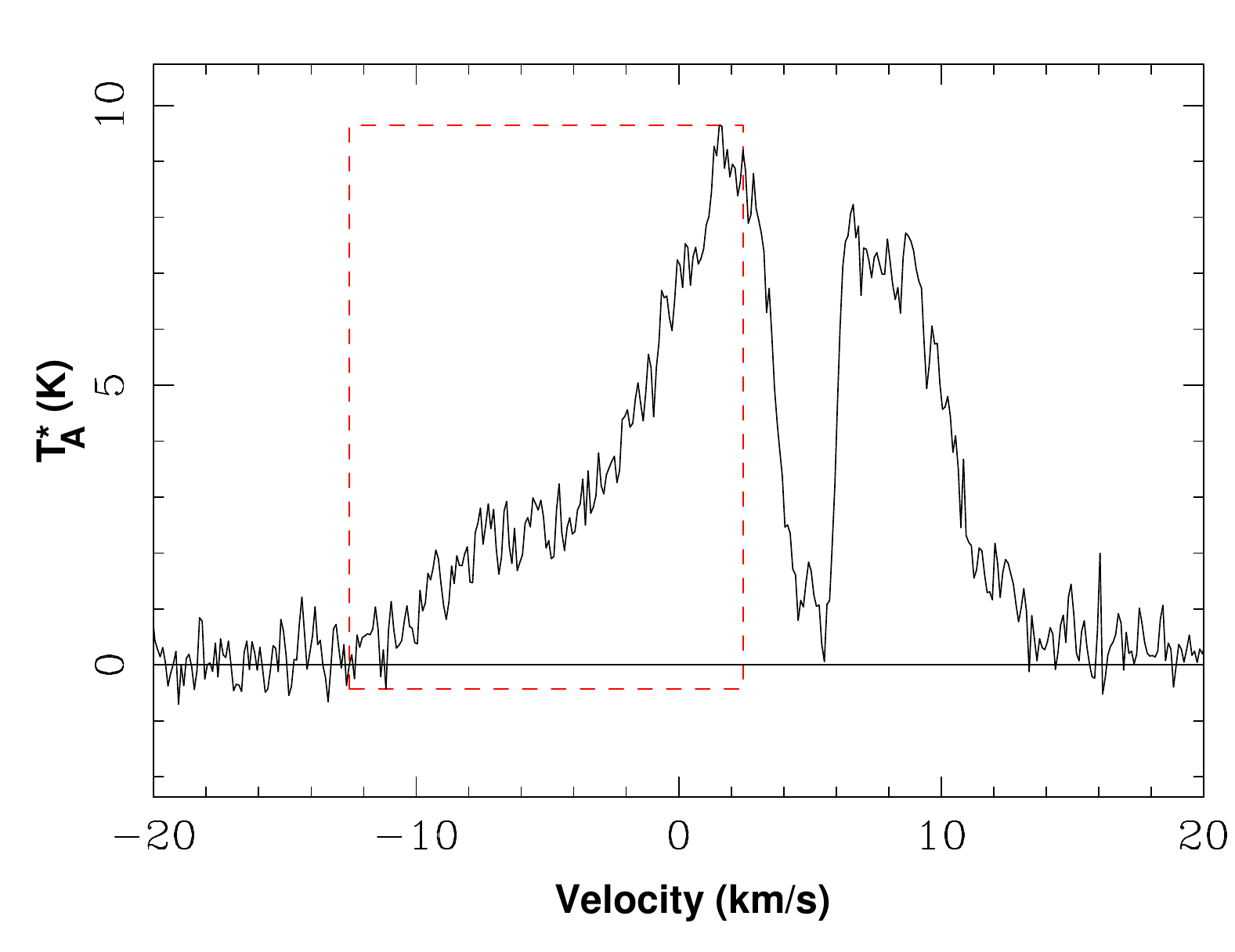}
  \includegraphics[width=0.5\textwidth]{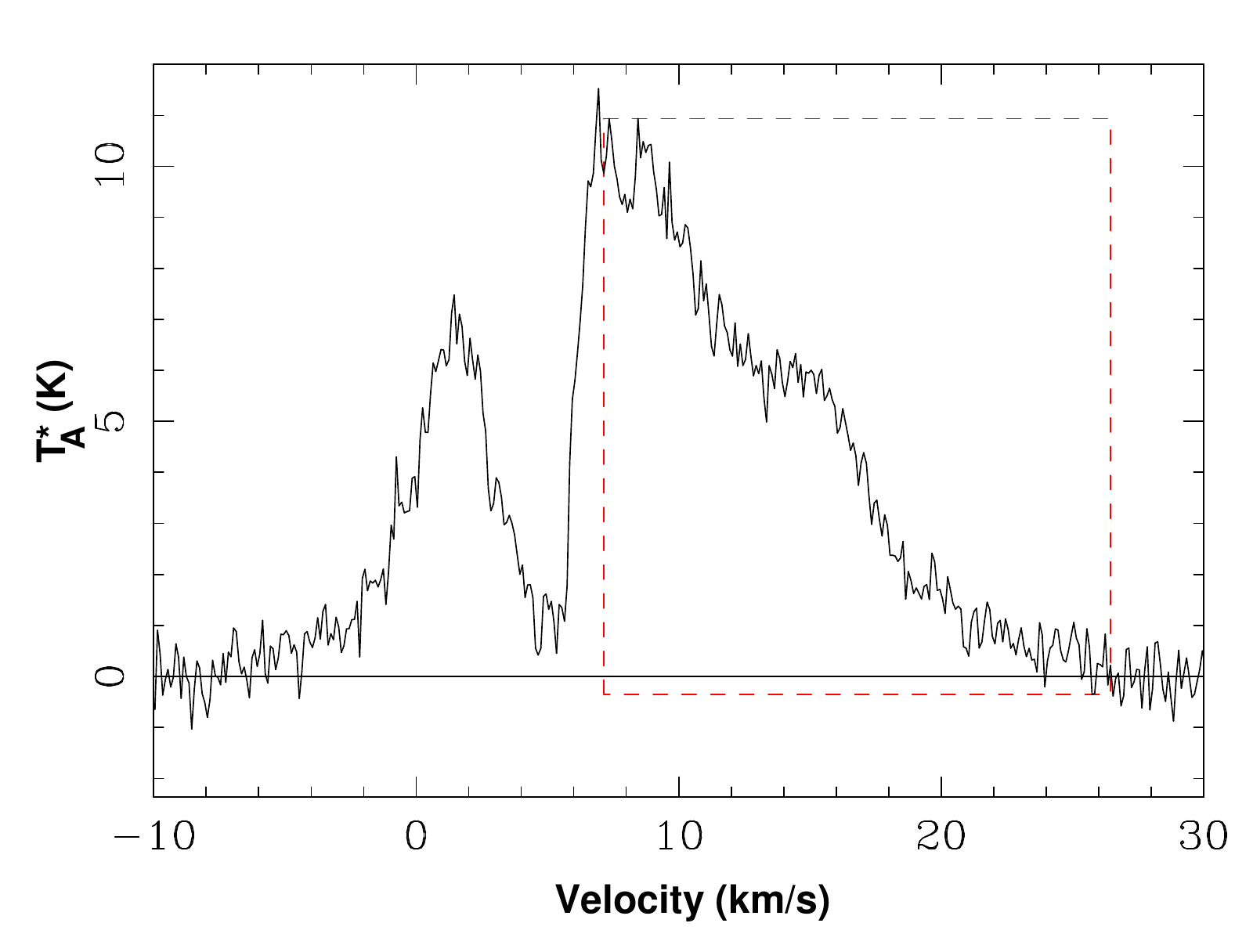}
  \caption{Up: $^{12}$CO J=3--2 spectrum with a blue-shifted wing obtained towards the position
represented by a blue square in Fig.\,\ref{figinteg}. Bottom: $^{12}$CO J=3--2 spectrum 
with a red-shifted wing obtained towards the position indicated with a red square in 
Fig.\,\ref{figinteg}. The dashed box in both spectra remarks the shifted emission components.}
  \label{spectra}
\end{figure}

From the $^{12}$CO data we look for kinematical spectral evidences of perturbed gas due to
the star-forming processes. Analyzing the whole data cube we found that in a region of about
1$^{\prime}$~centered at the position indicated with a blue square in Fig.\,\ref{figinteg} the
spectra present a blue-shifted wing, while in a region with the same size centered at the
red square (Fig.\,\ref{figinteg}) the spectra show a red-shifted wing. Two representative spectra
of both regions are presented in Fig.\,\ref{spectra}. These spectral features are strong evidences
of molecular outflows driven by the young stellar objects (YSOs) embedded in the cloud. A dynamically study
of the outflows will be presented in a further work.
Besides, the spectra present a dip at about 5 km s$^{-1}$, which shows that the emission is self-absorbed, 
suggesting the presence of high-density gas.

\begin{figure}[!ht]
  \centering
  \includegraphics[width=0.5\textwidth]{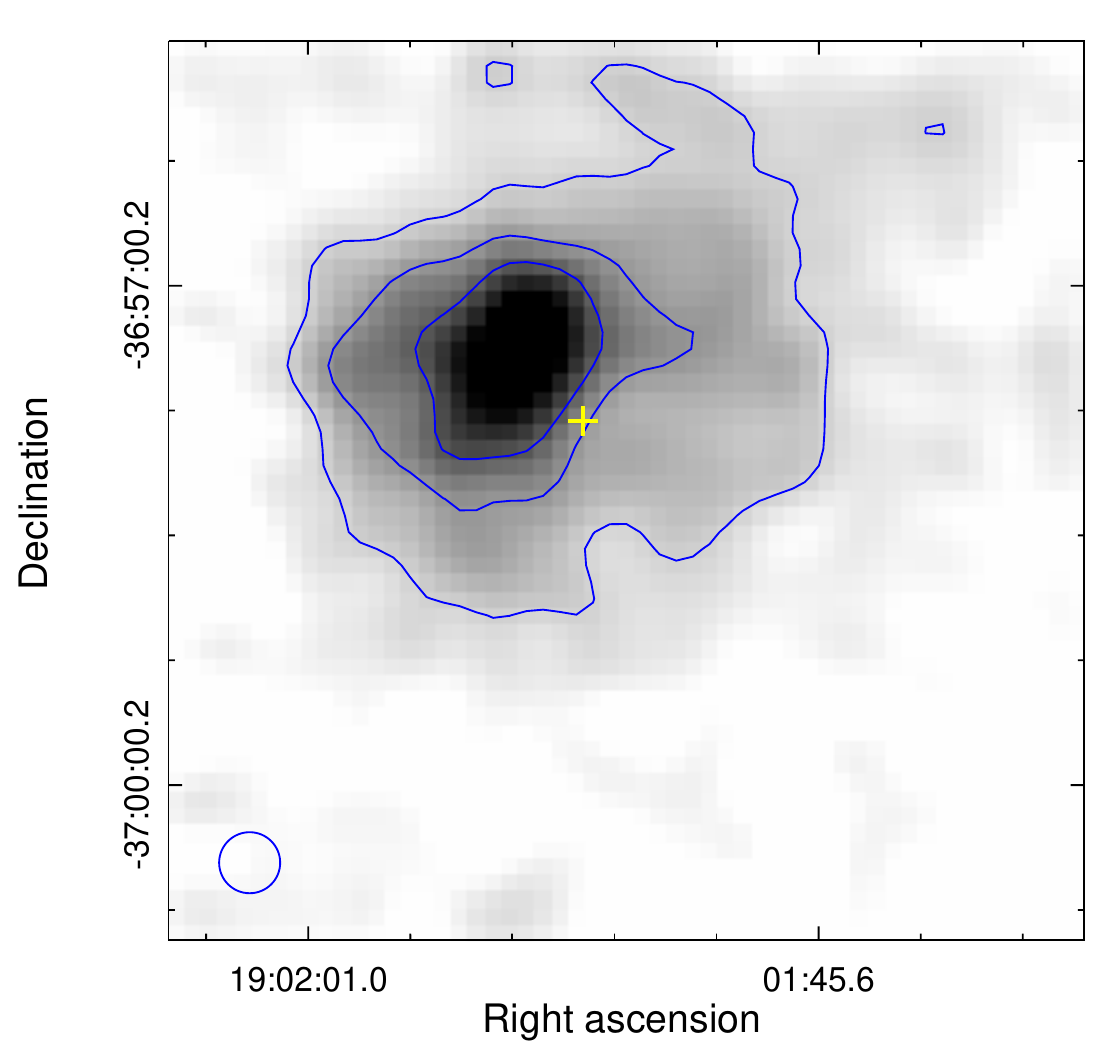}
  \includegraphics[width=0.48\textwidth]{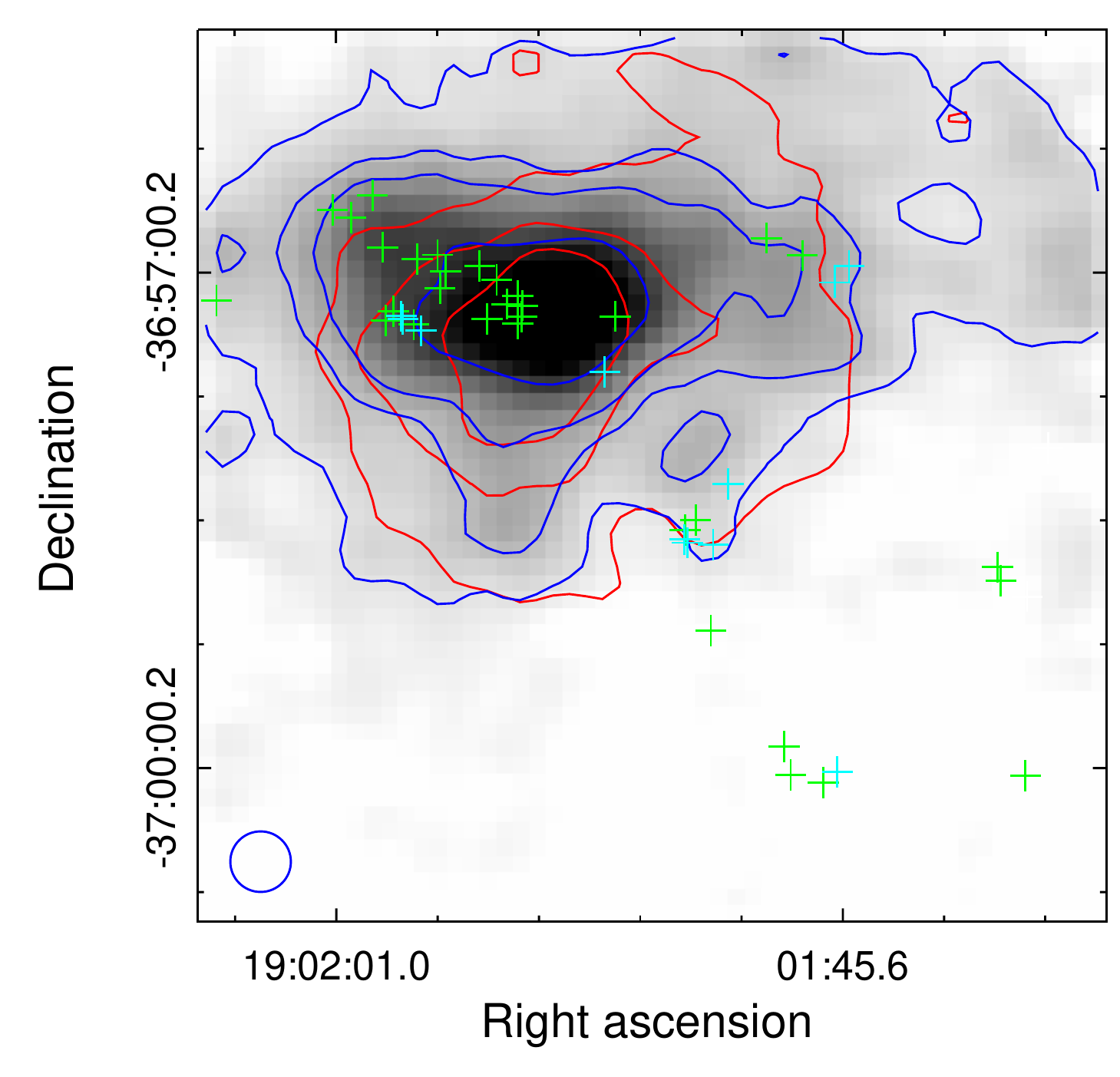}
  \caption{Up: HCO$^{+}$ J=4--3 emission integrated between $+2$ and $+10$ km~s$^{-1}$. The contour
levels are 5, 10, and 15 K km~s$^{-1}$. The beam is presented in the bottom-left corner. The yellow
cross is the position at which was observed the N$_{2}$H$^{+}$ J=4--3 line. Bottom: same as Fig.\,\ref{figinteg}
with the HCO$^{+}$ contours (in red) of the upper panel. }
  \label{fighco+}
\end{figure}

Figure \,\ref{fighco+} (up) displays the HCO$^{+}$ J=4--3 emission integrated between 
$+2$ and $+10$ km~s$^{-1}$, the range in which this line presents emission, 
towards the R CrA dark cloud. For comparison, Fig.\,\ref{fighco+} (bottom) presents the integrated 
$^{12}$CO emission and the stellar sources (as displayed in Fig.\,\ref{figinteg}) with the HCO$^{+}$ contours superimposed.  
It can be appreciated that the HCO$^{+}$ peak coincides with the $^{12}$CO maximum.  Most of the HHs and MHOs 
lie in the region delimited by the HCO$^{+}$ emission, suggesting that it
arises due to the outflowing phenomenon. It is known that such molecular species enhances
in molecular outflows \citep{raw04}. Indeed an enhancement of the HCO$^{+}$ abundance is
expected to occur due to the liberation and photoprocessing by shocks of the molecular material
stored in the icy mantles of the dust.

\begin{figure}[!ht]
  \centering
  \includegraphics[width=0.5\textwidth]{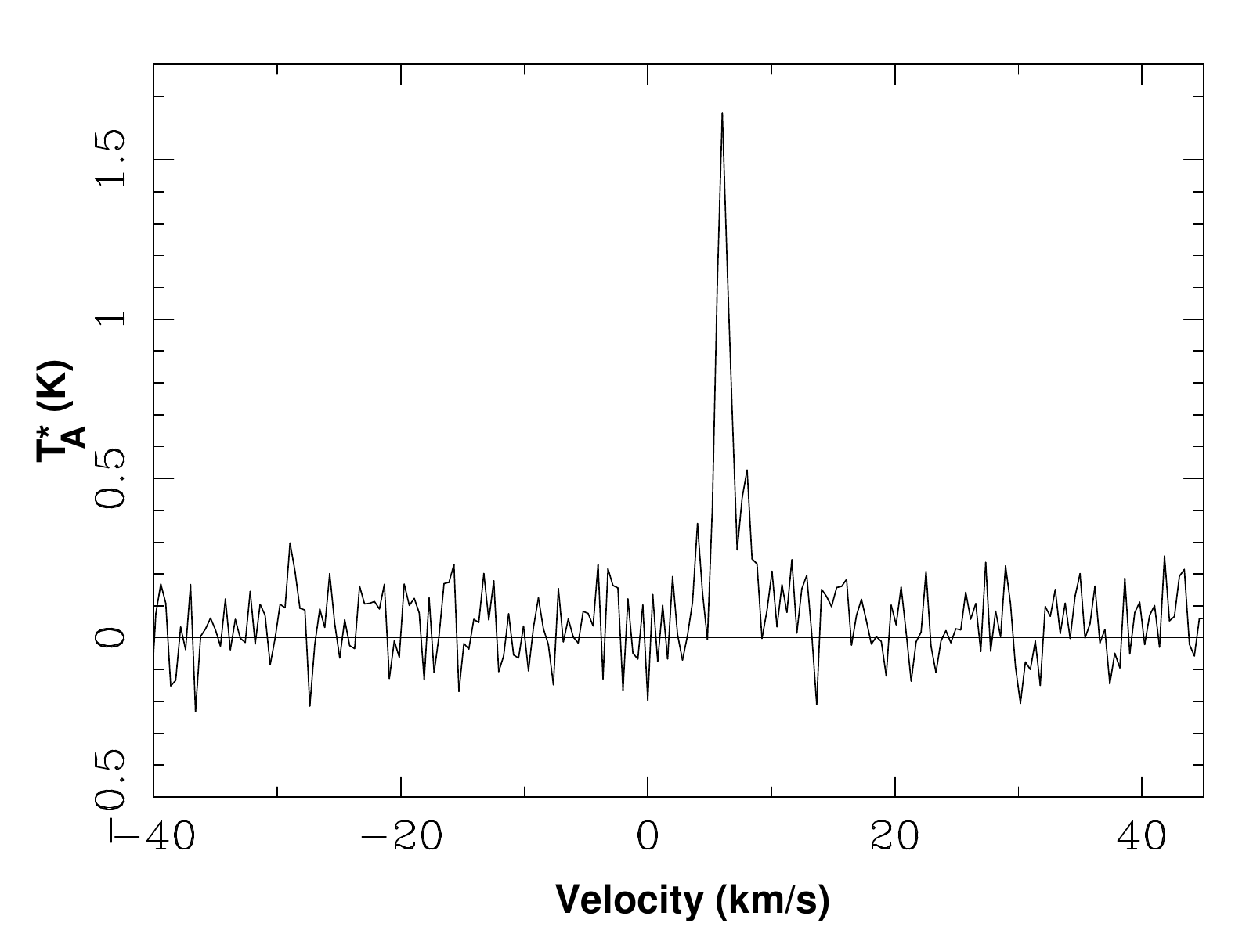}
  \caption{N$_{2}$H$^{+}$ J=4--3 emission obtained at 19:01:52.7, $-36$:57:49 (J2000) in
the R CrA dark cloud (yellow cross in Fig.\,\ref{fighco+}-up).}
  \label{fign2h+}
\end{figure}

Figure\,\ref{fign2h+} shows the spectrum of the N$_{2}$H$^{+}$ J=4--3 line obtained towards
the position indicated with a yellow cross in Fig.\,\ref{fighco+}. It is important to note
that this is the first reported detection of N$_{2}$H$^{+}$ in the J=4--3 line towards CrA dark cloud.
Its detection, centered at $v\sim6.12$ km s$^{-1}$, indicates the presence of high-density gas and it is an interesting line to study 
the ionization and chemistry related to the star-forming processes. For instance, it is believed 
that as star formation progresses, the temperature rises and thus the CO is released from the
dust, which then interacts with N$_{2}$H$^{+}$ through N$_{2}$H$^{+}$ $+$ CO $\rightarrow$ HCO$^{+}$ $+$ N$_{2}$
\citep{busquet11}.

Finally, we obtained a line integrated intensity ratio 
of $I(N_{2}H^{+})/I(HCO^{+}) \sim 0.33$ at the position in which the N$_{2}$H$^{+}$ was observed. This value
is close to the lower cutoff value assigned to the sources named as ``N$_{2}$H$^{+}$ poor'' in \citet{hoq13}.
Even though this study was performed for high-mass star forming regions and R CrA dark cloud hosts low and 
intermediate mass YSOs, we can suggest a similar chemistry for both types of YSOs. In this way, 
the YSOs related to the molecular gas analyzed here may be in the later stages of evolution \citep{hoq13}.

\begin{acknowledgement}
The ASTE project is driven by Nobeyama Radio Observatory (NRO), a branch
of National Astronomical Observatory of Japan (NAOJ), in collaboration
with University of Chile, and Japanese institutes including University of
Tokyo, Nagoya University, Osaka Prefecture University, Ibaraki University,
Hokkaido University and Joetsu University of Education.
S.P. and M.O. are members of the {\sl Carrera del
investigador cient\'\i fico} of CONICET, Argentina. M.C.P. is a doctoral
fellow of CONICET, Argentina.
This work was partially supported by Argentina grants awarded by UBA (UBACyT), CONICET and ANPCYT.
M.R. wishes to acknowledge support from CONICYT through FONDECYT grant N$^{\rm o}$ 1140839.

\end{acknowledgement}


\bibliographystyle{baaa}
\small
\bibliography{biblio}

\begin{thebibliography}{}

\bibitem[\protect\citeauthoryear{{Busquet}, {Estalella}, {Zhang}, {Viti},
  {Palau}, {Ho} \& {S{\'a}nchez-Monge}}{{Busquet} et~al.}{2011}]{busquet11}
{Busquet} G.,    et~al., 2011, \aap, 525, A141

\bibitem[\protect\citeauthoryear{{Davis}, {Gell}, {Khanzadyan}, {Smith} \&
  {Jenness}}{{Davis} et~al.}{2010}]{davis10}
{Davis} C.~J.,    et~al., 2010, \aap, 511, A24

\bibitem[\protect\citeauthoryear{{Hoq}, {Jackson}, {Foster}, {Sanhueza},
  {Guzm{\'a}n}, {Whitaker}, {Claysmith}, {Rathborne}, {Vasyunina} \&
  {Vasyunin}}{{Hoq} et~al.}{2013}]{hoq13}
{Hoq} S.,    et~al., 2013, \apj, 777, 157

\bibitem[\protect\citeauthoryear{{Lindberg} \& {J{\o}rgensen}}{{Lindberg} \&
  {J{\o}rgensen}}{2012}]{linde12}
{Lindberg} J.~E.,  {J{\o}rgensen} J.~K.,  2012, \aap, 548, A24

\bibitem[\protect\citeauthoryear{{Neuh{\"a}user} \& {Forbrich}}{{Neuh{\"a}user}
  \& {Forbrich}}{2008}]{neuha08}
{Neuh{\"a}user} R.,  {Forbrich} J.,  2008, {The Corona Australis Star Forming
  Region in Handbook of Star Forming Regions, Volume II, Ed. B. Reipurth}.
p.~735

\bibitem[\protect\citeauthoryear{{Rawlings}, {Redman}, {Keto} \&
  {Williams}}{{Rawlings} et~al.}{2004}]{raw04}
{Rawlings} J.~M.~C.,    et~al., 2004, \mnras, 351, 1054

\bibitem[\protect\citeauthoryear{{Reipurth}}{{Reipurth}}{2000}]{reip00}
{Reipurth} B.,  2000, VizieR Online Data Catalog, 5104, 0

\end{thebibliography}
 
\end{document}